\documentclass[aps,pre,twocolumn,showpacs,preprintnumbers,amsmath,amssymb]{revtex4}
\usepackage{epsfig}

\newcommand{\compresslist}{%
\setlength{\itemsep}{0pt}%
\setlength{\parskip}{0pt}%
\setlength{\parsep}{0pt}%
}

\begin{document}
\title{Analytical approach to model of scientific revolutions}
\author{Pawe{\l} Kondratiuk, Grzegorz Siudem and Janusz A. Ho{\l}yst}
\affiliation{Faculty of Physics, Center of Excellence for Complex Systems Research, Warsaw University of Technology, Koszykowa 75, PL-00-662 Warsaw, Poland}
\date{\today}

\begin{abstract}
The model of scientific paradigms spreading throughout the community of agents with memory is analyzed using the master equation. The case of two competing ideas is considered for various networks of interactions, including agents placed at Erd\H{o}s-R\'{e}nyi graphs or complete graphs. The pace of adopting a new idea by a community is analyzed, along with the distribution of periods after which a new idea replaces the old one. The approach is extended for the chain topology onto the more general case when more than two ideas compete. Our analytical results are in agreement with numerical simulations.
\end{abstract}

\maketitle


\section{INTRODUCTION}
There is a tendency to separate certain periods in the history of civilizations, such as Renaissance or Enlightenment, which qualitatively differ from each other by dominating trends in science, art or customs. Technological innovations and scientific discoveries constantly emerge and it is not likely that some kind of equilibrium --- "end of history" \cite{Fukuyama92} --- will ever be reached. The changes (evolutionary and revolutionary ones) happen due to the   interactions and exchange of innovative ideas \cite{Ebeling00,Coleman57, Axelrod97, Castellano09} at the level of individuals, communities or even civilizations. Eventually, ideas spread throughout the communities \cite{Bettencourt06, Ratkiewicz10}. Some of the ideas gain broad (even global) acceptance and popularity, replacing old ones \cite{Krapivsky11}. Similar phenomena can be observed in models of opinion formation dynamics \cite{Castellano09,Kacperski96,Kacperski99,Schweitzer00,Lorenz07}.

The process of adoption of an innovative technology \cite{Frenken11} or a new scientific concept by individuals and communities differs from adoption of, for example, a new trend in arts. Obsolete technologies and discarded scientific theories, once abandoned, are not likely to be accepted by individuals again. To model such a process, agents should be given some kind of memory. Another important fact is that the will of individuals to adopt a new scientific concept depends on its global popularity. For example, spreading of technological innovations is often slowed down by incompatibility with existing standards. In the field of arts, the situation is different. Old ideas can reemerge and become popular again, such as Renaissance artists were inspired by Antique philosophy or architecture.

Recently a model was introduced by Bornholdt et al. which attempts to describe scientific revolutions \cite{Bornholdt11}. The model combines interactions at the level of individuals with influence of the whole community. Despite its simplicity, it managed to reconstruct some key features of the dynamics of scientific paradigms spreading, including an asymmetry between the rate of adopting a new idea by the community and the speed of its decline when new competing ideas appear. The model was based on numerical simulations and no analytical treatment was presented. In this paper, the master equation and Markov processes theory \cite{Gardiner85} were applied to analyze the dynamics of the system in the case of small level of agents' creativity for various topologies of agent interactions, such as chain, complete graph, ER graphs, star and square lattice. For chain topology the approach was extended in an attempt to describe the system dynamics for higher levels of creativity.


\section{THE MODEL OF SPREADING OF IDEAS}
\label{sec:model}
The rules of the model \cite{Bornholdt11} are very simple. $N$ agents occupy nodes of a network. Every agent follows some paradigm (idea), labeled by a natural number. In each time step a random agent $i$ (with paradigm $s_i$) is selected, along with one of its neighbors $j$ (with paradigm $s_j$). If the agent $i$ has never followed the paradigm $s_j$, it adopts it with probability $N_{s_j}/N$, where $N_{s_j}$ denotes the number of agents representing paradigm $s_j$. Additionally, new paradigms, which have never been present in the community, can appear: with probability $\alpha$ a random agent is selected, which changes its paradigm into one which has never been present in the community.

The most important feature of the model is the {\it memory} of the agents, who do not adopt the same paradigm twice. One can find analogy between this model and evolutionary dynamics models: innovations can be regarded as mutations which allow the affected individuals outperform their rivals. Lack of any evident {\it fitness} parameter, which would describe how well a specie is adapted to the environment, is not necessarily a drawback of such an interpretation, as the {\it fitness} of a specie is alway {\it a posteriori} knowledge \cite{Klimek10}.

There are some general features of the evolution of the system, independent of the interactions network topology. For very small probability $\alpha$ at most two paradigms coexist (other cases are neglectable due to their much smaller probability). This case will be analyzed for various networks in Sec. \ref{sec:very_small_alpha}.

For higher values of probability $\alpha$, other effects have to be considered. In such a case usually more than two paradigms coexist, which ``compete'' with each other. However, one may suppose that within a relatively wide range of $\alpha$ still two paradigms can be separated at every moment: the ``old'' paradigm which is the most popular, but currently at the decline, and the ``new'' paradigm, the second most popular, which will prevail after some time (and then enter the stage of decline). This case will be analyzed in Sec. \ref{sec:small_alpha}.


\section{THE CASE OF TWO COMPETING PARADIGMS}
\label{sec:very_small_alpha}

\subsection{General case}
When the creativity level of the agents $\alpha$ is small enough, at most 2 paradigms coexist, referred to as the paradigm 0 (at the stage of decline) and the paradigm 1 (at the stage of expansion). The evolution consists of two distinct periods.
\begin{enumerate}
\compresslist
\item All the agents share the same paradigm 0. The length of this {\it stage of stagnation} is a random variable of the exponential distribution
\begin{equation}
P(T_{stag}) = \alpha (1-\alpha)^{T_{stag}} 
\label{eqn:Tstag_dist}
\end{equation}
and the mean value
\begin{eqnarray}
\langle T_{stag} \rangle & = & \sum_{T_{stag} = 0}^{\infty} T_{stag} \alpha (1-\alpha)^{T_{stag}}  \nonumber \\ 
 & = & \frac{1-\alpha}{\alpha} \approx \frac{1}{\alpha}.
\end{eqnarray}
\item After an innovative paradigm 1 appears, it starts spreading across the community. The {\it time of expansion} of the paradigm 1 will be denoted $T$. It is a random variable whose distribution depends on the interactions network topology. After time $T$ all the agents share the paradigm 1 and the state of the system is equivalent to the initial one.
\end{enumerate}

In our approach the state of the system is characterized by one variable --- the number $n$ of agents sharing paradigm 1. The problem reduces to the problem of expansion of the paradigm 1 throughout the community, starting from one agent with the paradigm 1 at time $t=0$. The generic master equation has only two terms:
\begin{equation}
\frac{\partial}{\partial t} P(n, t) = P(n-1, t) W_{n,n-1} - P(n, t) W_{n+1,n},
\label{eqn:generic_master}
\end{equation}
where transition rates from the state $n$ to $n+1$ (for $n \in [1, N-1]$) are equal to
\begin{equation}
W_{n+1,n}  \equiv W_n = \frac{n}{N^2} \sum_{i=1}^N \frac{1-s_i}{k_i} \sum_{j=1}^N a_{ij} s_j
\end{equation}
where $s_i$ denotes the state of $i\text{th}$ agent ($s_i = 0 \Rightarrow$ agent follows the old paradigm, $s_i = 1 \Rightarrow$ agent follows the new paradigm), $k_i$ is the degree of node $i$ and $a_{ij}$ is the adjacency matrix. For $n = N$ so defined transition rate is automatically equal to 0, since $\forall_i s_i = 0$ then.

It can be easily proved that if all the transition rates are different ($k \neq j \Rightarrow W_k \neq W_j$), the solution of Eq. \eqref{eqn:generic_master} with the initial condition $P(n,0) = \delta_{n1}$ is
\begin{equation}
P(n,t) = \sum_{k=1}^{n} C_k^n e^{-W_k t},
\label{eqn:generic_P}
\end{equation}
where
\begin{equation}
C_k^n \equiv \prod_{i=1}^{n-1} W_i \prod_{\substack{j=1 \\ j \neq k}}^{n} \frac{1}{W_j - W_k}.
\end{equation}

Let us note that since $W_N = 0$ and $\forall_{1\leq n<N} W_n > 0$, the distribution evolves into
\begin{equation}
\lim_{t \rightarrow \infty} P(n,t) = \delta_{nN}
\end{equation}
(all the agents share the paradigm 1), which is an expected limit.

The approximation of only two competing paradigms makes sense if the mean {\it stagnation time} $\langle T_{stag} \rangle = 1/\alpha$ is greater than the mean {\it expansion time} $\langle T \rangle$, i.e.
\begin{equation}
\alpha < \frac{1}{\langle T \rangle}.
\label{eqn:alpha_limit}
\end{equation}
This upper limit of $\alpha$ has to be estimated for each type of network separately. In general the $P(T)$ distribution can be expressed by the $P(n,t)$ probability as
\begin{eqnarray}
P(T=t) = W_{N-1} P(N-1,t-1) \approx \frac{1}{N} P(N-1,t)
\label{eqn:generic_T_dist}
\end{eqnarray}
and, taking into account Eq. \eqref{eqn:generic_P}, we obtain
\begin{eqnarray}
\langle T \rangle & \approx & \int_0^\infty t P(T=t) dt \approx \frac{1}{N} \sum_{k=1}^{N-1} \frac{C_{k}^{N-1}}{W_{k}^{2}}.
\label{eqn:generic_T_avg}
\end{eqnarray}

\subsection{Chain topology}

Let us consider the case when the agents occupy nodes of a chain. For simplicity, periodic boundary conditions will be assumed.

This specific topology makes the problem quite simple and in the first approximation it is not necessary to analyze the master equation \eqref{eqn:generic_master}. The average number of agents sharing paradigm 1 can be derived from the recursive equation
\begin{equation}
  \left\{
  \begin{array}{l l}
    \langle n(0) \rangle = 1 \\
    \langle n(t+1) \rangle = \langle n(t) \rangle + \frac{2}{N} \frac{1}{2} \frac{\langle n(t) \rangle}{N} = \langle n(t) \rangle \left(1 + \frac{1}{N^2} \right),
  \end{array} \right.
\label{eqn:n_avg_0_rec}
\end{equation}
which has the solution
\begin{equation}
	\langle n(t) \rangle = \left(1 + \frac{1}{N^2} \right)^t \approx e^{t/N^2}.
\label{eqn:n_avg_0}
\end{equation}
On the average, after time
\begin{equation}
\langle T\rangle = N^2 \log N
\label{eqn:chain_T_avg_approx}
\end{equation}
the paradigm 1 will stop spreading as it will be shared by the whole community. The situation will be stable until another innovation appears. From this condition the range of $\alpha$ can be estimated for which this approximation makes sense. Taking into account Eq. \eqref{eqn:alpha_limit} we get
\begin{equation}
	\alpha < \frac{1}{N^2 \log N}.
\end{equation}

For a more exact analysis one has to consider the master equation \eqref{eqn:generic_master}. Let us make a simple observation that the subgraph consisting of agents sharing the new paradigm is connected. The transition rates are therefore equal to
\begin{eqnarray}
W_n & = & \frac{n}{N^2} \sum_{i=1}^N \frac{1-s_i}{2} (s_{i+1} + s_{i-1}) \nonumber \\
& = & \frac{n}{N^2} (1-\delta_{nN}).
\label{eqn:chain_W}
\end{eqnarray}
Above we used the periodic boundary conditions, so agents at positions $1$ and $N+1$ are equivalent.

To solve the problem, at first we will neglect the $\delta_{nN}$ term and treat the $n$ variable as if it could grow to infinity, $n=1,2,\dots,\infty$. Eventually, the transition rates from state $n$ to $n+1$ are equal to $W_n = \frac{n}{N^2}$ and the master equation has the form
\begin{equation}
\frac{\partial}{\partial t} P(n, t) = P(n-1, t) \frac{n-1}{N^2} - P(n, t) \frac{n}{N^2}.
\label{eqn:chain_master}
\end{equation}

Due to the simple form of the transition rates, Eq. \eqref{eqn:chain_master} can be solved using the method of characteristic function $G$:
\begin{equation}
G(s,t) \equiv \langle e^{ins} \rangle.
\end{equation}
This approach has such an advantage over using \eqref{eqn:generic_P}, that solutions are automatically in a compact form. The master equation \eqref{eqn:chain_master} with the initial condition $P(n,0) = \delta_{n0}$ leads to the partial differential equation with the initial condition
\begin{equation}
\left\{\begin{array}{l l}
    \frac{\partial}{\partial t} G(s,t) + \frac{1}{iN^2} \left(e^{is} - 1 \right) \frac{\partial}{\partial s} G(s,t) = 0 \\
    G(s,0) = e^{is},
  \end{array} \right.
\end{equation}
which can be solved as
\begin{equation}
G(s,t) = \frac{1}{1-e^{t/N^2}(1-e^{-is})}.
\end{equation}
After a short algebra it can be proved that
\begin{equation}
G(s,t) = \sum_{n=1}^{\infty} \frac{1}{e^{t/N^2}-1} \left(1-e^{-t/N^2}\right)^n e^{isn},
\end{equation}
so
\begin{equation}
P(n,t) = e^{-t/N^2} \left(1-e^{-t/N^2}\right)^{n-1}.
\end{equation}
This is valid for $n<N$. In order to take into consideration the limitation on the $n$ variable (the $\delta_{nN}$ term in Eq. \eqref{eqn:chain_W}), one has to consider the accumulation of probability at point $n=N$:
\begin{eqnarray}
P(n=N,t) & = & \sum_{m=N}^{\infty} e^{-t/N^2} \left(1-e^{-t/N^2}\right)^{m-1} \nonumber \\
 & = & \left(1-e^{-t/N^2}\right)^{N-1}.
\end{eqnarray}
Eventually,
\begin{equation}
  P(n,t) = \left\{\begin{array}{l l l}
    e^{-t/N^2} \left(1-e^{-t/N^2}\right)^{n-1} &, 1 \leq n < N \\
    \left(1-e^{-t/N^2}\right)^{N-1} &, n=N.
  \end{array} \right.
\label{eqn:chain_P}
\end{equation}
The mean value of $n$ resulting from the distribution \eqref{eqn:chain_P} is equal to
\begin{equation}
\langle n \rangle = e^{t/N^2} \left( 1 - \left( 1 - e^{-t/N^2} \right)^N \right),
\label{eqn:chain_mean}
\end{equation}
which for small $t$ reduces to \eqref{eqn:n_avg_0}.

\begin{figure}
\centering
\includegraphics[scale=0.7, angle=270]{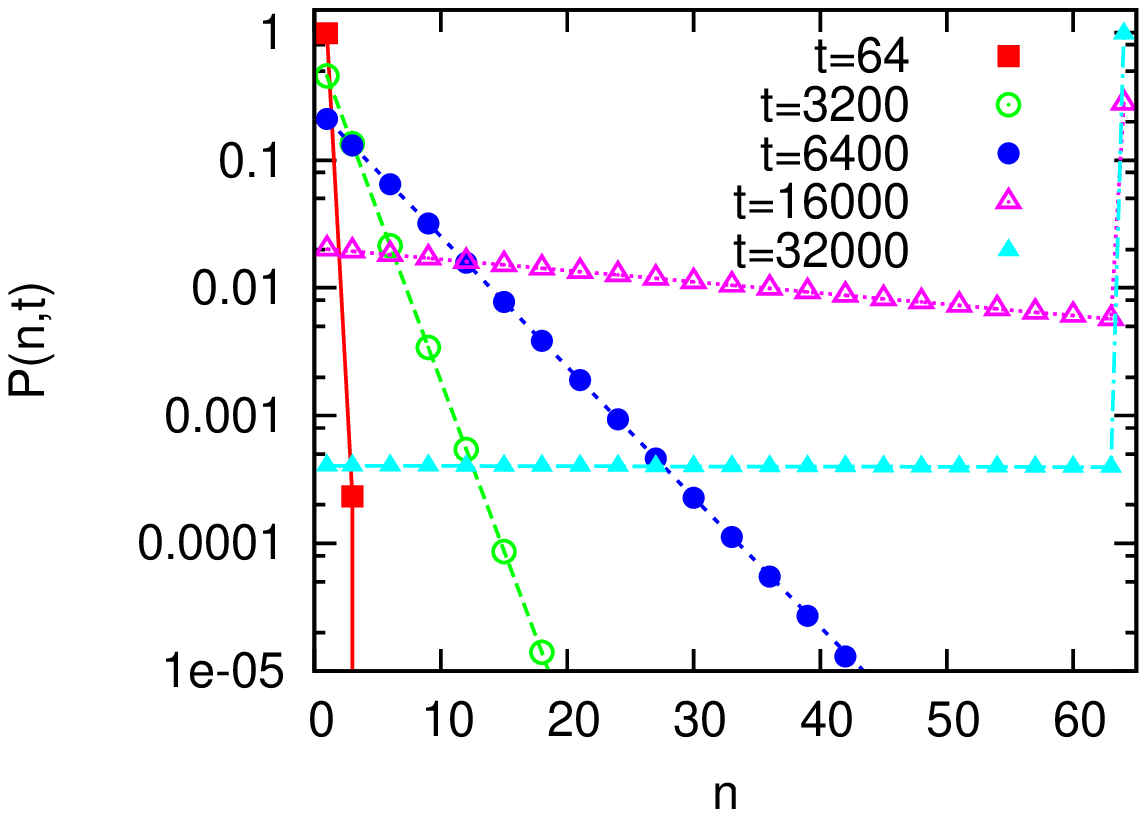}
\caption{(Color online) Chain graph topology, $N=64$ nodes, $\alpha < 1/\langle T \rangle$. Evolution of the system starting from $P(n,t=0)=\delta_{n1}$; probability $P(n,t)$ at various moments $t$. Points are obtained from the numerical solution of the master equation \eqref{eqn:chain_master}. Lines --- analytical predictions (Eq. \eqref{eqn:chain_P}).}
\label{fig:chain_dist_evol}
\end{figure}

\begin{figure}
\centering
\includegraphics[scale=0.7, angle=270]{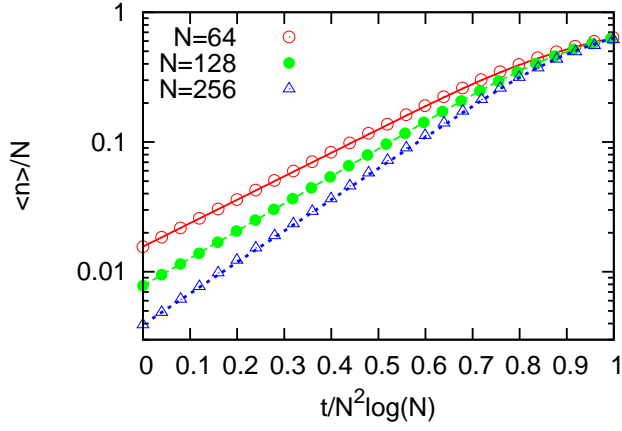}
\caption{(Color online) Chain graph topology, $N=64,128,256$ nodes, $\alpha < 1/\langle T \rangle$. Evolution of the system starting from $n=1$ innovative agent: $\langle n \rangle$ versus time. Points --- simulated data. Lines --- analytical predictions (Eq. \eqref{eqn:chain_mean}).}
\label{fig:chain_means}
\end{figure}

From the distribution $P(n,t)$, a more exact approximation of $\langle T \rangle$ than Eq. \eqref{eqn:chain_T_avg_approx} can be obtained. According to Eq. \eqref{eqn:generic_T_dist},
\begin{equation}
P(T=t) \approx \frac{1}{N} e^{-t/N^2} \left(1-e^{-t/N^2}\right)^{N-2},
\label{eqn:chain_T_dist}
\end{equation}
and
\begin{eqnarray}
\langle T \rangle & = & \sum_{t=0}^{\infty} t P(T=t) \nonumber \\
& \approx & \frac{1}{N} \int_{0}^{\infty} t e^{-t/N^2} \left(1-e^{-t/N^2}\right)^{N-2} dt \nonumber\\
& = & N^2 H_{N-1},
\label{eqn:chain_T_avg_exact}
\end{eqnarray}
where $H_n$ is the $n$th harmonic number. Since harmonic numbers grow approximately as fast as the natural logarithm, the approximated solution \eqref{eqn:chain_T_avg_approx} is very close to \eqref{eqn:chain_T_avg_exact}. In fact, for $N \gg 1$
\begin{equation}
H_{N} \approx \log N + \gamma,
\end{equation}
where $\gamma \approx 0.5772$ denotes the Euler-Mascheroni constant. 

\begin{figure}
\centering
\includegraphics[scale=0.7, angle=270]{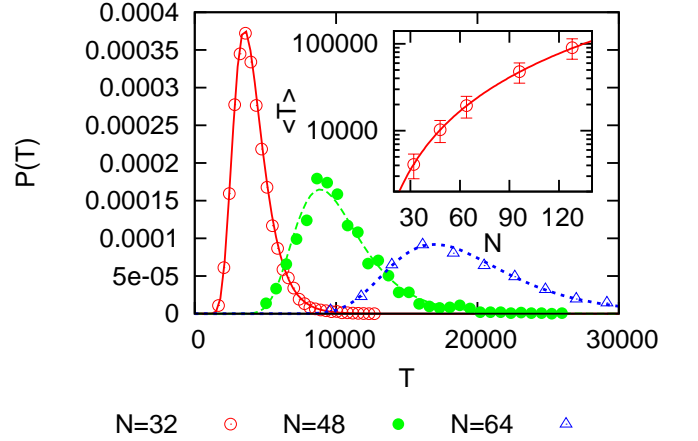}
\caption{(Color online) Chain graph topology, $\alpha < 1/\langle T \rangle$. Distribution of the expansion periods lengths $T$ for different system sizes $N$. Points --- simulations, lines --- analytical predictions (Eq. \eqref{eqn:chain_T_dist}). (Inset) Mean time of expansion $\langle T\rangle$ versus system size $N$: simulations (points with error bars) compared with the analytical predictions (line) (Eq. \eqref{eqn:chain_T_avg_exact}).}
\label{fig:chain_T_dist}
\end{figure}


\subsection{Complete graph topology}

Let us consider the situation when interaction is possible between every pair of agents, i.e. $\forall_{(i,j)} a_{ij} = 1$. Referring to the generic master equation \eqref{eqn:generic_master}, the transition rates are equal to
\begin{equation}
W_n = \frac{n}{N^2} \sum_{i=1}^N \frac{1-s_i}{N-1} \sum_{j=1}^N s_j = \frac{n^2(N-1)}{N^2(N-1)} \approx \frac{n^2(N-1)}{N^3}
\label{eqn:compl_W}
\end{equation}
Eventually, the master equation has the form
\begin{eqnarray}
\frac{\partial}{\partial t} P(n,t) & = & P(n-1,t)\frac{(n-1)^2(N-n+1)}{N^3} \nonumber \\
& - & P(n,t) \frac{n^2(N-n)}{N^3}.
\label{eqn:compl_master}
\end{eqnarray}
If all the transition rates are different ($j \neq k \Rightarrow W_j \neq W_k$, which is satisfied if equation $N = a(1 + b + b^2)$ does not have trivial solutions $a,b$ among natural numbers), the solution can be written in the form of the sum \eqref{eqn:generic_P}:
\begin{equation}
P(n,t) = \sum_{k=1}^{n} C_k^n e^{-k^2(N-k) t/N^3},
\label{eqn:compl_P}
\end{equation}
where
\begin{eqnarray}
C_k^n & \equiv & \prod_{i=1}^{n-1} i^2(N-i) \prod_{\substack{j=1 \\ j \neq k}}^{n} \frac{1}{j^2(N-j) - k^2(N-k)} \nonumber \\
& = & (n-1)!^2 \frac{(N-1)!}{(N-n)!} \prod_{\substack{j=1 \\ j \neq k}}^{n} \frac{1}{j^2(N-j) - k^2(N-k)}
\end{eqnarray}

The expansion time $T$ distribution $P(T)$ can be derived from the $P(n,t)$ distribution:
\begin{equation}
P(T=t) \approx \frac{1}{N} P(N-1,t) = \frac{1}{N} \sum_{k=1}^{N-1} C_k^{N-1} e^{-k^2(N-k) t/N^3}.
\label{eqn:compl_T_dist}
\end{equation}

\begin{figure}
\centering
\includegraphics[scale=0.7, angle=270]{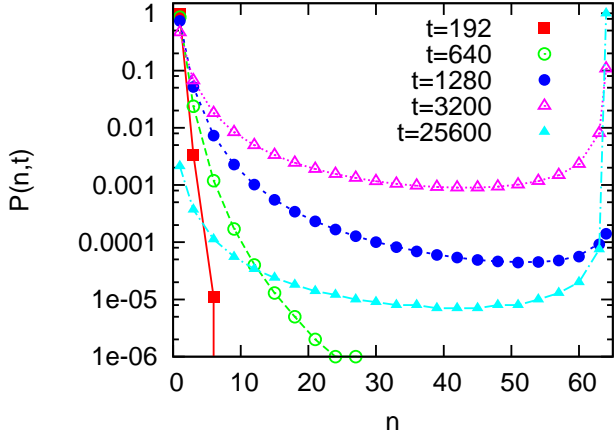}
\caption{(Color online) Complete graph topology, $N=64$ nodes, $\alpha < 1/\langle T \rangle$. Evolution of the system starting from $P(n,t=0)=\delta_{n1}$; probability $P(n,t)$ at various moments $t$. Points are obtained from the numerical solution of the master equation \eqref{eqn:compl_master}. Lines --- analytical predictions (Eq. \eqref{eqn:compl_P}).}
\label{fig:compl_dist_evol}
\end{figure}

\begin{figure}
\centering
\includegraphics[scale=0.7, angle=270]{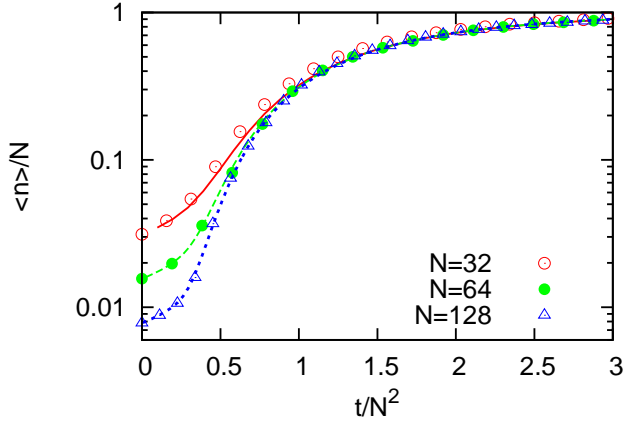}
\caption{(Color online) Complete graph topology, $N=32,64,128$ nodes, $\alpha < 1/\langle T \rangle$. Evolution of the system starting from $n=1$ innovative agent: $\langle n \rangle$ versus time. Points --- simulated data. Lines --- analytical predictions (obtained from Eq. \eqref{eqn:compl_P}).}
\label{fig:compl_means}
\end{figure}

\begin{figure}
\centering
\includegraphics[scale=0.7, angle=270]{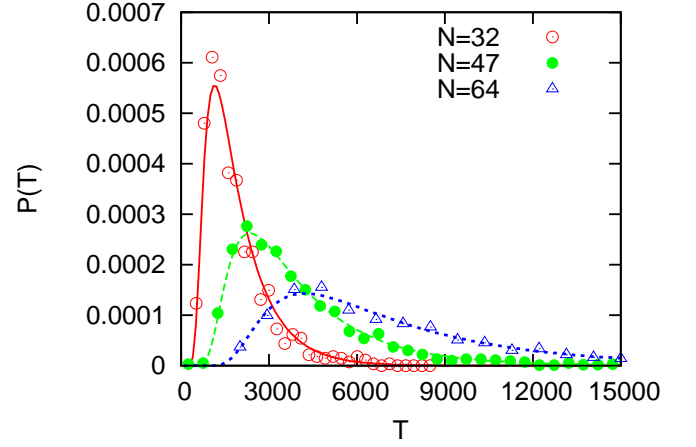}
\caption{(Color online) Complete graph topology, $\alpha < 1/\langle T \rangle$. Distribution of the expansion period lengths $T$ for different system sizes $N$. Points --- simulations, lines --- analytical predictions (Eq. \eqref{eqn:compl_T_dist}).}
\label{fig:compl_T_dist}
\end{figure}

The analytical predictions are in agreement with the simulations (Fig. \ref{fig:compl_dist_evol}--\ref{fig:compl_T_dist}).

\subsection{Erd\H{o}s-R\'{e}nyi graph topology}

Let us consider the situation when the network of interactions is an Erd\H{o}s-R\'{e}nyi graph \cite{Albert02}. For each pair of nodes the edge between them exists with probability $p$.

The degree distribution of an ER graph is a binomial distribution
\begin{equation}
P(k) = Bin(N,p) \equiv \binom{N-1}{k} p^{k} (1-p)^{N-1-k}
\end{equation}

The transition rates of the master equation \eqref{eqn:generic_master} are equal to
\begin{equation}
W_n = \frac{n}{N^2} \sum_{i=1}^N (1-s_i) \sum_{j=1}^N \frac{a_{ij} s_j}{k_i} = \frac{n}{N^2} \sum_{i=1}^N (1-s_i) \frac{k_i^+}{k_i^+ + k_i^-},
\end{equation}
where the random variable $k_i^+$ denotes number of $i$'s neighbors following the paradigm 1 and $k_i^-$ --- the paradigm 0. Within the mean field approach, the transition rates can be estimated by
\begin{equation}
W_n = \frac{n(N-n)}{N^2} \left \langle \frac{k^+}{k^+ + k^-} \right \rangle,
\label{eqn:er_W}
\end{equation}
where $\langle \cdot \rangle$ denotes averaging on the whole population of the agents. Treating $k^+$ and $k^-$ as independent random variables with binomial distributions, we obtain
\begin{eqnarray}
\left \langle \frac{k^+}{k^+ + k^-} \right \rangle & = & \sum_{k^+=1}^{n} \sum_{k^-=0}^{N-n-1} \frac{k^+}{k^+ + k^-} P(k^+) P(k^-) \nonumber \\
& = & \sum_{k^+=1}^{n} k^+ \binom{n}{k^+} p^{k^+} (1-p)^{n-k^+} \nonumber \\
& & \cdot \sum_{k^-=0}^{N-n-1} \frac{\binom{N-n-1}{k^-}}{k^+ + k^-}  p^{k^-} (1-p)^{N-n-1-k^-} \nonumber \\
& = & (1-p)^{N-n-1} \sum_{k^+=1}^{n} k^+ \binom{n}{k^+} p^{k^+} (1-p)^{n-k^+} \nonumber \\
& & \cdot \left(\frac{1-p}{p}\right)^{k^+} \int\limits_{0}^{p/(1-p)} \xi^{k^+-1} (1+\xi)^{N-n-1} d \xi \nonumber \\
& = & \frac{n}{N-1} \left( 1 - (1-p)^{N-1} \right).
\end{eqnarray}
Eventually,
\begin{eqnarray}
W_n  = \frac{n^2(N-n)}{N^3} \left( 1 - (1-p)^{N-1} \right).
\end{eqnarray}
For $p=1$ the transition rates reduce, as expected, to the ones obtained for the complete graph topology.

The master equation has the form
\begin{eqnarray}
\frac{\partial}{\partial t} P(n,t) & = & \left( P(n-1,t)\frac{(n-1)^2(N-n+1)}{N^3} \right. \\ \nonumber
& - & \left. P(n,t) \frac{n^2(N-n)}{N^3} \right) \left( 1 - (1-p)^{N-1} \right).
\label{eqn:er_master}
\end{eqnarray}
Similarly to the case of complete graph topology, the solution can be written in the form of the sum \eqref{eqn:generic_P}:
\begin{equation}
P(n,t) = \sum_{k=1}^{n} C_k^n e^{-k^2(N-k) \left( 1 - (1-p)^{N-1} \right) t/N^3},
\label{eqn:er_P}
\end{equation}
where
\begin{eqnarray}
C_k^n & \equiv & \prod_{i=1}^{n-1} i^2(N-i) \prod_{\substack{j=1 \\ j \neq k}}^{n} \frac{1}{j^2(N-j) - k^2(N-k)} \nonumber \\
& = & (n-1)!^2 \frac{(N-1)!}{(N-n)!} \prod_{\substack{j=1 \\ j \neq k}}^{n} \frac{1}{j^2(N-j) - k^2(N-k)}
\end{eqnarray}

Now the {\it expansion time} distribution $P(T)$ is
\begin{eqnarray}
P(T=t) & \approx & \frac{1}{N} P(N-1,t) \nonumber \\
& = & \frac{1}{N} \sum_{k=1}^{N-1} C_k^{N-1} e^{-k^2(N-k) \left( 1 - (1-p)^{N-1} \right) t/N^3}.
\label{eqn:er_T_dist}
\end{eqnarray}

\begin{figure}
\centering
\includegraphics[scale=0.7, angle=270]{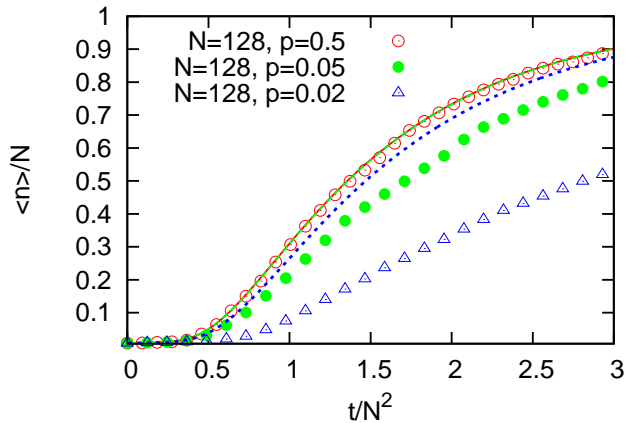}
\caption{(Color online) ER graph topology, $N=128$ nodes, $p=0.5,0.05,0.02$, $\alpha < 1/\langle T \rangle$. Evolution of the system starting from $n=1$ innovative agent: $\langle n \rangle$ versus time. Points --- simulated data. Lines --- analytical predictions (obtained from Eq. \eqref{eqn:er_P}).}
\label{fig:er_means}
\end{figure}

As can be seen in Fig. \ref{fig:er_means}, our approach predicts a decline in the rate of growth of the new paradigm cluster with decreasing the network density (parameter $p$), but seriously underestimates that decline. We suppose that there are some nontrivial correlations between the agents' states resulting from the dynamics, which were not taken into consideration.


\subsection{Star topology}

\subsubsection{Central agent innovative}

Let us consider the star topology of interactions, i.e. there exists a {\it central} agent connected to all the other $N-1$ {\it periferal} agents --- and those are the only connections. Moreover, we will require that at time $t=0$ the central agent follows the innovative paradigm 1. This case is interesting since in this variant the new idea spreads at the highest pace --- the transition rates
\begin{equation}
W_n = \frac{(N-n)n}{N^2}
\label{eqn:st0_W}
\end{equation}
are the highest possible for this model among all the possible topologies (one can for example compare Eq. \eqref{eqn:st0_W} to Eq. \eqref{eqn:chain_W}, \eqref{eqn:compl_W}, \eqref{eqn:er_W}, \eqref{eqn:sq_W}).

The periferal agents are not connected with each other and they are only influenced by the ``mean field'' of paradigms. Therefore, one can consider the change in time of the average state of a periferal agent. Let $\pi(t)$ denote probability that at time $t$ a periferal agent follows the paradigm 1. The evolution of $\pi(t)$ follows the recursive equation
\begin{equation}
  \left\{\begin{array}{c c l}
    \pi(0) & = & 0 \\
    \pi(t+1) & = & \pi(t) + (1-\pi(t)) \frac{1}{N-1} \frac{\langle n(t) \rangle}{N} \\
    & = & \pi(t) + \frac{1}{N} (1-\pi(t)) \left(\pi(t) + \frac{1}{N-1} \right),
  \end{array} \right.
\label{eqn:st0_pi_rec}
\end{equation}
which can be solved in the approximation of continuous time:
\begin{equation}
\frac{d \pi}{d t} \approx - \frac{1}{N} (\pi - 1) \left (\pi + \frac{1}{N-1} \right)
\end{equation}
\begin{equation}
\pi(t) = \frac{1 + \frac{1}{N-1}}{(N-1) \exp(-\frac{t}{N-1}) + 1} - \frac{1}{N-1}.
\end{equation}
Thus, the average number of agents sharing the paradigm 1 is equal to
\begin{equation}
\langle n(t) \rangle = (N-1)\pi(t) + 1 \approx \frac{N}{1 + N e^{-t/N}}.
\label{eqn:st0_avg_n}
\end{equation}

\begin{figure}
\centering
\includegraphics[scale=0.7, angle=270]{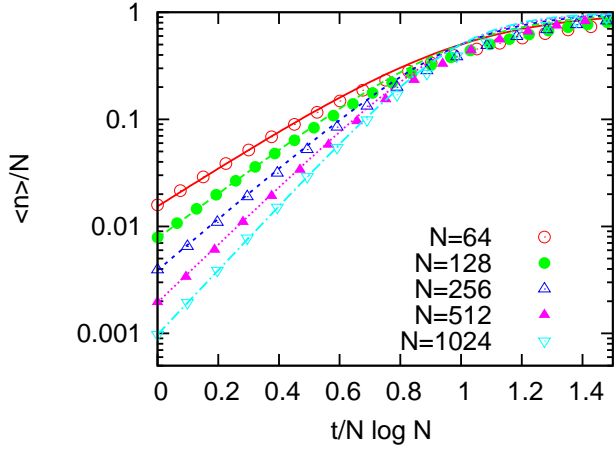}
\caption{(Color online) Star lattice topology with central agent innovative, $N=64,128,256,512,1024$ nodes, $\alpha < 1/\langle T \rangle$. Evolution of the system starting from $n=1$ innovative agent (central): $\langle n \rangle$ versus time. Points --- simulated data. Lines --- analytical predictions (Eq. \eqref{eqn:st0_avg_n})}
\label{fig:st0_means}
\end{figure}

From $\pi(t)$, the {\it expansion time} distribution $P(T)$ can be derived:
\begin{eqnarray}
P(T = t) & \approx & \frac{d}{dt} P(T \leq t) = \frac{d}{dt} (\pi(t))^{N-1} \nonumber \\
& \approx & \frac{N e^{t/N} (e^{t/N} - 1)^{N-2}}{(e^{t/N} + N - 1)^{N}}.
\label{eqn:st0_T_dist}
\end{eqnarray}

\begin{figure}
\centering
\includegraphics[scale=0.7, angle=270]{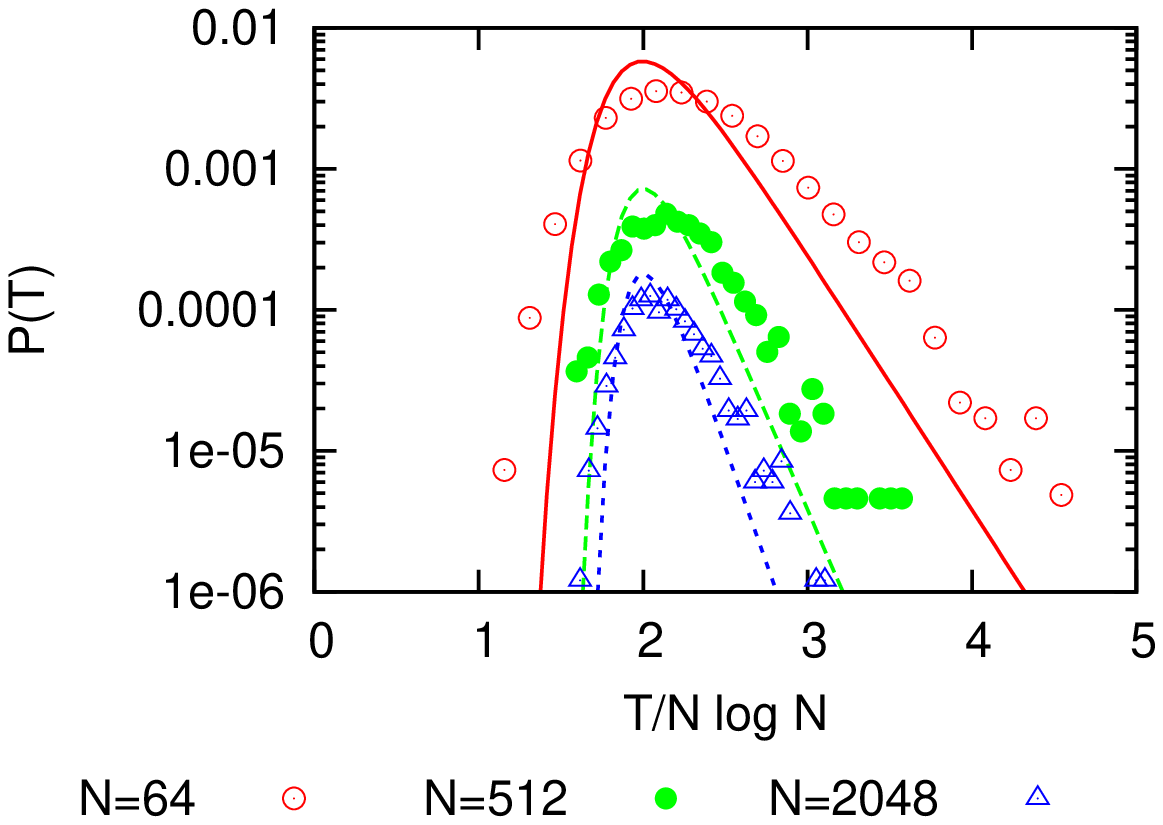}
\caption{(Color online) Star topology with central agent innovative, $\alpha < 1/\langle T \rangle$. Distribution of the expansion period lengths  $T$ for different system sizes $N$. Points --- simulations, lines --- analytical predictions (Eq. \eqref{eqn:st0_T_dist}).}
\label{fig:st0_T_dist}
\end{figure}

We were not able to find an analytical formula for $\langle T \rangle$. However, since $P(T)$ is unimodal with a well defined maximum, one can assume that $\langle T \rangle$ grows as fast with $N$ as
\begin{eqnarray}
\arg\max P(T) & = & N \log \left( \frac{\sqrt{N^4 - 2 N^3 + N^2 - 4 N + 4}}{2} \right. \nonumber \\
& + & \left. \frac{N^2 - N}{2}\right) = O(N \log N)
\end{eqnarray}
Indeed, as can be seen in Fig. \ref{fig:st_avgT}, the mean {\it expansion time} can be very well approximated by
\begin{eqnarray}
\langle T \rangle \approx k N \log N,
\label{eqn:st0_avgT}
\end{eqnarray}
where parameter $k$, as obtained by fitting to simulated data, is equal to $k = 2.149 \pm 0.007$. Since the transition rates \eqref{eqn:st0_W} are the highest possible, the mean {\it expansion time} \eqref{eqn:st0_avgT} must be the shortest for this model among all the possible topologies.

\subsubsection{All agents equally innovative}

The system is similar to the one described above. The only difference is that at time $t=0$ any agent can be the innovative one. With probability $(N-1)/N \approx 1$ a periferal agent will become the innovative one and after time $T_0$, which is a random variable with exponential probability distribution
\begin{eqnarray}
P(T_0 = t) & = & \left(1 - \frac{1}{N^2(N-1)} \right)^{t-1} \frac{1}{N^2(N-1)} \nonumber \\
& \approx & \frac{1}{N^3} e^{-t/N^3},
\end{eqnarray}
the central agent will adopt idea 1. Then the dynamics will be as described in the case of innovative central agent. Thus, the probability distribution of the {\it expansion time} can be well approximated by a convolution of two probability distributions:
\begin{eqnarray}
P(T=t) & \approx & \int_{0}^{\infty} \frac{N e^{\tau/N} (e^{\tau/N} - 1)^{N-2}}{(e^{\tau/N} + N - 1)^{N}} \frac{1}{N^3} e^{(-t + \tau)/N^3} d \tau \nonumber \\
& \approx & \frac{1}{N^3} e^{-t/N^3} = P(T_0 = t).
\label{eqn:st1_T_dist}
\end{eqnarray}

The mean and the standard deviation of this exponential probability distribution are equal to
\begin{equation}
\langle T \rangle = \sigma(T) = N^3.
\label{eqn:st1_avgT}
\end{equation}

\begin{figure}
\centering
\includegraphics[scale=0.7, angle=270]{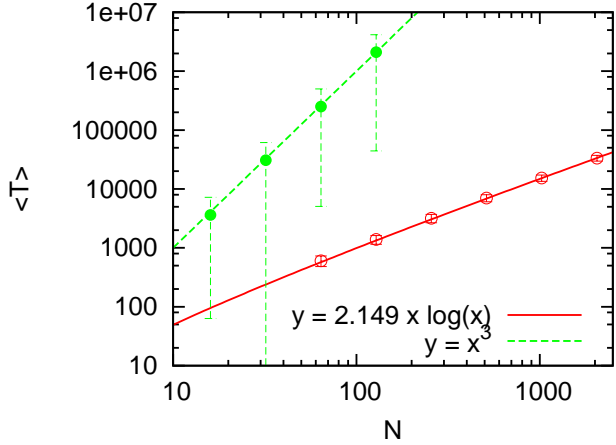}
\caption{(Color online) Star lattice topology. Mean time of expansion $\langle T\rangle$ versus system size $N$. Empty circles --- innovative central agent, filled circles --- all agents equally innovative. Error bars correspond to the standard deviations of the samples. Lines --- analytical predictions, Eq. \eqref{eqn:st0_avgT} and \eqref{eqn:st1_avgT}.}
\label{fig:st_avgT}
\end{figure}

As can be seen in Fig. \ref{fig:st_avgT}, the analytical results for both variants of star topology agree with simulations very well.

\begin{figure}
\centering
\includegraphics[scale=0.7, angle=270]{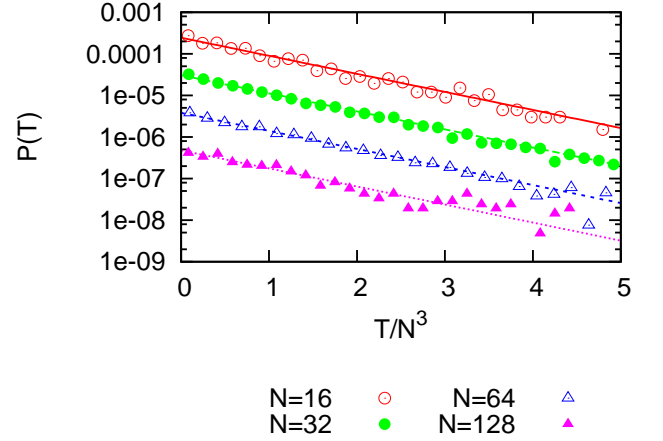}
\caption{(Color online) Star topology with all agents equally innovative, $\alpha < 1/\langle T \rangle$. Distribution of the expansion period lengths $T$ for different system sizes $N$. Points --- simulations, lines --- analytical predictions (Eq. \eqref{eqn:st1_T_dist}).}
\label{fig:st1_T_dist}
\end{figure}

\subsection{Square lattice topology}

Let us consider the square lattice topology. Periodic boundary conditions will be assumed, so each agent has 4 neighbors. The first approximation would be to assume that the cluster of agents sharing paradigm 1 grows uniformly in each direction, so at any moment it is circle-shaped, with the radius of the circle equal to $r = \sqrt{n/\pi}$. Therefore, within this approximation the transition rates in the generic master equation \eqref{eqn:generic_master} are equal to
\begin{equation}
W_n = \frac{\sqrt{\pi} n^{3/2}}{2 N^2} (1-\delta_{nN}).
\label{eqn:sq_W}
\end{equation}

\begin{figure}
\centering
\includegraphics[scale=0.7, angle=270]{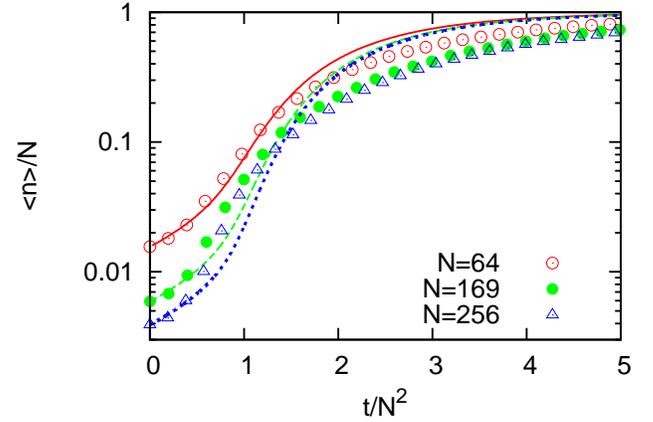}
\caption{(Color online) Square lattice topology, $N=64,169,256$ nodes, $\alpha < 1/\langle T \rangle$. Evolution of the system starting from $n=1$ innovative agent: $\langle n \rangle$ versus time. Points --- simulated data. Lines --- analytical predictions (obtained from Eq. \eqref{eqn:generic_P} with transition rates \eqref{eqn:sq_W}).}
\label{fig:sq_means}
\end{figure}

Similarly as in the case of complete graph topology, the solution of the master equation
\begin{eqnarray}
\frac{\partial}{\partial t} P(n,t) & = & P(n-1,t) \frac{\sqrt{\pi} (n-1)^{3/2}}{2 N^2} (1-\delta_{n-1,N}) \nonumber \\ 
& - & P(n,t) \frac{\sqrt{\pi} n^{3/2}}{2 N^2} (1-\delta_{nN}).
\label{eqn:square_master}
\end{eqnarray}
can be expressed in the form of the sum of products \eqref{eqn:generic_P}. Having compared the results of such an approximation with the simulations (Fig. \ref{fig:sq_means}), one has to state that this approach significantly overestimates the pace of the growth of the new paradigm cluster.

\section{THE CASE OF MANY COMPETING IDEAS}
\label{sec:small_alpha}

If the mean {\it stagnation time} $\langle T_{stag}\rangle = 1/\alpha$ is shorter than the mean {\it expansion time} $\langle T \rangle$, i.e.
\begin{equation}
\alpha = \frac{1}{\langle T_{stag}\rangle}  > \frac{1}{\langle T \rangle},
\end{equation}
it is most probable that more than two paradigms coexist in the community at any moment. This case is much more difficult to describe analytically. Below we present our results for chain topology, which is probably the simplest one.


For $\alpha$ higher than $1/{N^2 \log N}$ Eq. \eqref{eqn:n_avg_0_rec} has to be extended by terms describing the appearing of new clusters of ideas, which would slower the process of expansion of the paradigm 1. In the first approximation it will be assumed that the only important new clusters are those appearing inside the cluster of the paradigm 1 and that they do not overlap with each other. Their growth is described by Eq. \eqref{eqn:n_avg_0}. Thus, the recursive equation for $\langle n(t) \rangle$ is now
\begin{eqnarray}
    \langle n(t+1) \rangle & = & \langle n(t) \rangle \left(1 + \frac{1}{N^2} \right) \nonumber \\
    & - & \sum_{\tau=0}^{t} \alpha \frac{\langle n(\tau)\rangle}{N} \langle\Delta n_{new}(t-\tau)\rangle \nonumber\\
    & = &  \langle n(t) \rangle \left(1 + \frac{1}{N^2} \right) \nonumber \\ 
    & - & \frac{\alpha}{N^3} \sum_{\tau=0}^{t} \langle n(\tau)\rangle \exp\left(\frac{t-\tau}{N^2}\right).
\label{eqn:n_avg_1_rec}
\end{eqnarray}
Substituting the sum with the integral and stating that $\langle n(t) \rangle = \exp\left(t/N^2\right)f(t)$ leads to the following equation for $f(t)$:
\begin{equation}
f'(t) + \frac{\alpha}{N^3} \int_0^t f(\tau) d\tau = 0,
\end{equation}
which, assuming the same initial conditions as in Eq. \eqref{eqn:n_avg_0_rec} (single innovation at time $t=0$), has the solution
\begin{equation}
f(t) = \cos\left(\lambda t\right),
\end{equation}
where $\lambda \equiv \sqrt{\alpha/N^3}$. The complete formula for the first approximation of $\langle n(t)\rangle$ is therefore
\begin{equation}
\langle n(t) \rangle = \exp\left(\frac{t}{N^2}\right) \cos\left(\lambda t\right).
\label{eqn:n_avg_1}
\end{equation}
As it should have been expected, for $\alpha \to 0$ this approximation converges to the previous one (Eq. \eqref{eqn:n_avg_0}).

A better approximation can be obtained by substituting the term $\exp\left(\frac{t-\tau}{N^2}\right)$ by $\langle n(t-\tau) \rangle$ in Eq. \eqref{eqn:n_avg_1_rec}, as new paradigms can also be "attacked" by paradigms appearing after them. The equation
\begin{eqnarray}
    \langle n(t+1) \rangle & = &  \langle n(t) \rangle \left(1 + \frac{1}{N^2} \right) \\ \nonumber
    & - & \frac{\alpha}{N^3} \sum_{\tau=0}^{t} \langle n(\tau)\rangle \langle n(t-\tau) \rangle
\label{eqn:n_avg_2_rec}
\end{eqnarray}
does not have a simple analytical solution, but by substituting the sum with the integral and stating $\langle n(t) \rangle = \exp\left(t/N^2\right)\cos(\lambda t)(1+g(t))$, where $g(t) \ll 1$, an integral equation can be obtained,
\begin{eqnarray}
0 & = & -\lambda \sin(\lambda t) +  g'(t) \cos(\lambda t) \\
& + & \lambda^2 \int_0^t \cos(\lambda \tau) (1+g(\tau)) \nonumber \\
&  & \cdot \cos(\lambda (t-\tau)) (1+g(t-\tau)) d\tau, \nonumber
\end{eqnarray}
which can be is solved provided all the terms in the integral apart from the product $\cos(\lambda \tau) \cos(\lambda(t-\tau))$ are neglected. Eventually, the second approximation of $\langle n(t) \rangle$ obtains the form
\begin{equation}
\langle n(t) \rangle = \exp\left(\frac{t}{N^2}\right) \cos(\lambda t) (1-\log|\cos(\lambda t)| - \frac{1}{4} \lambda^2 t^2).
\label{eqn:n_avg_2}
\end{equation}

\begin{figure*}
\centering
\begin{tabular}{ccc}
\includegraphics[width=0.22\textwidth, angle=270]{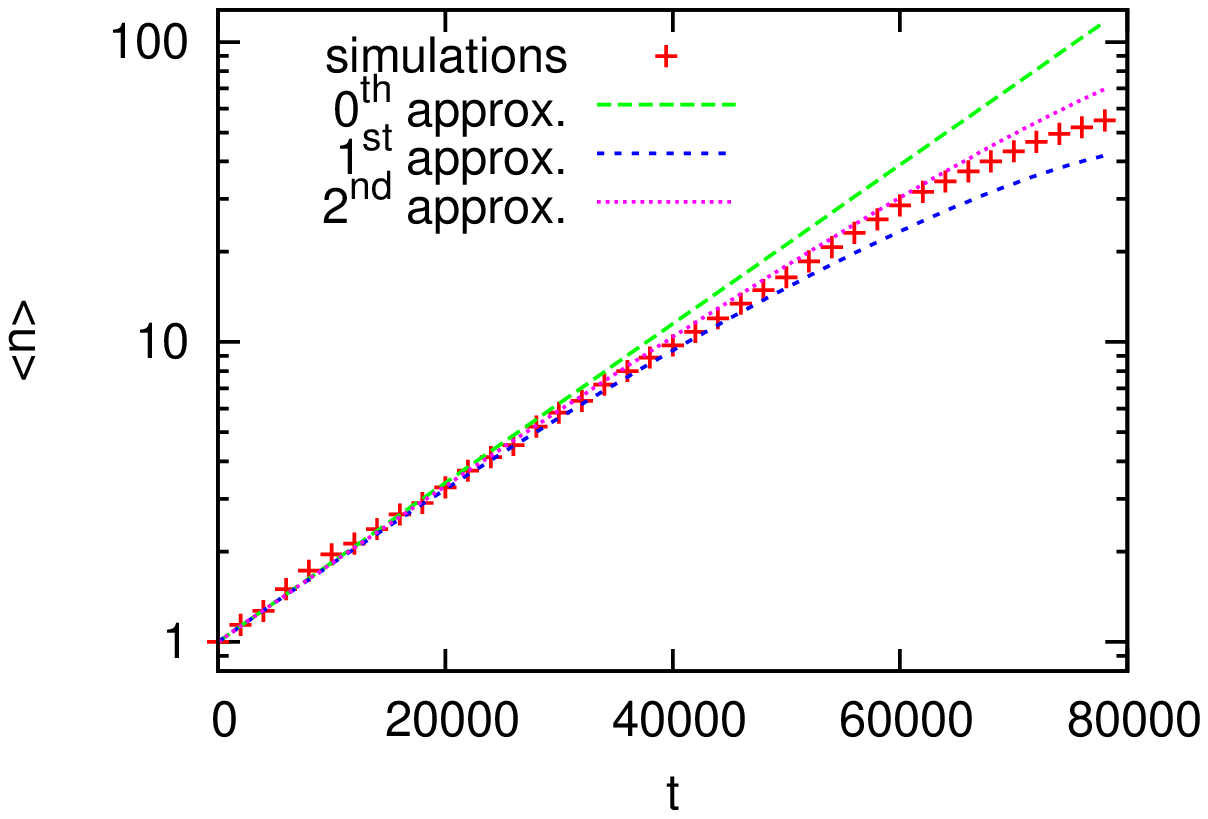} & 
\includegraphics[width=0.22\textwidth, angle=270]{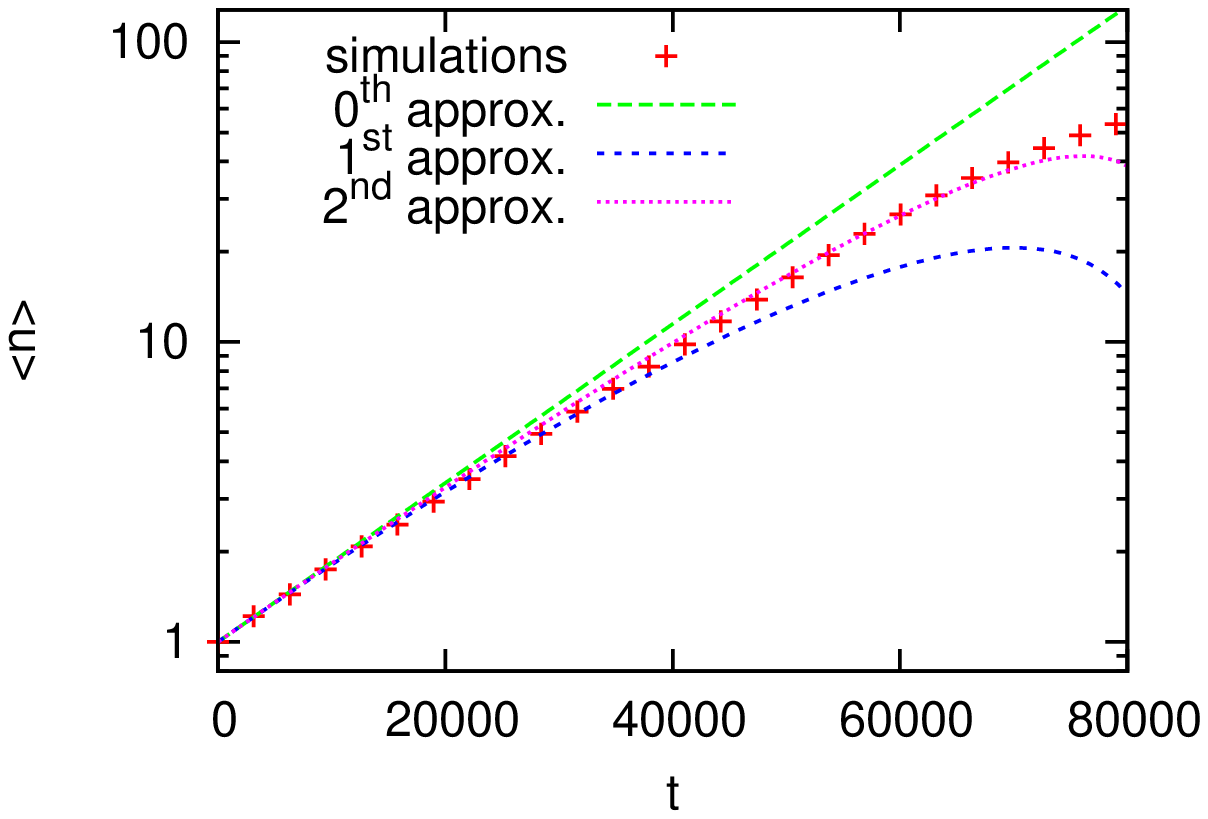} & 
\includegraphics[width=0.22\textwidth, angle=270]{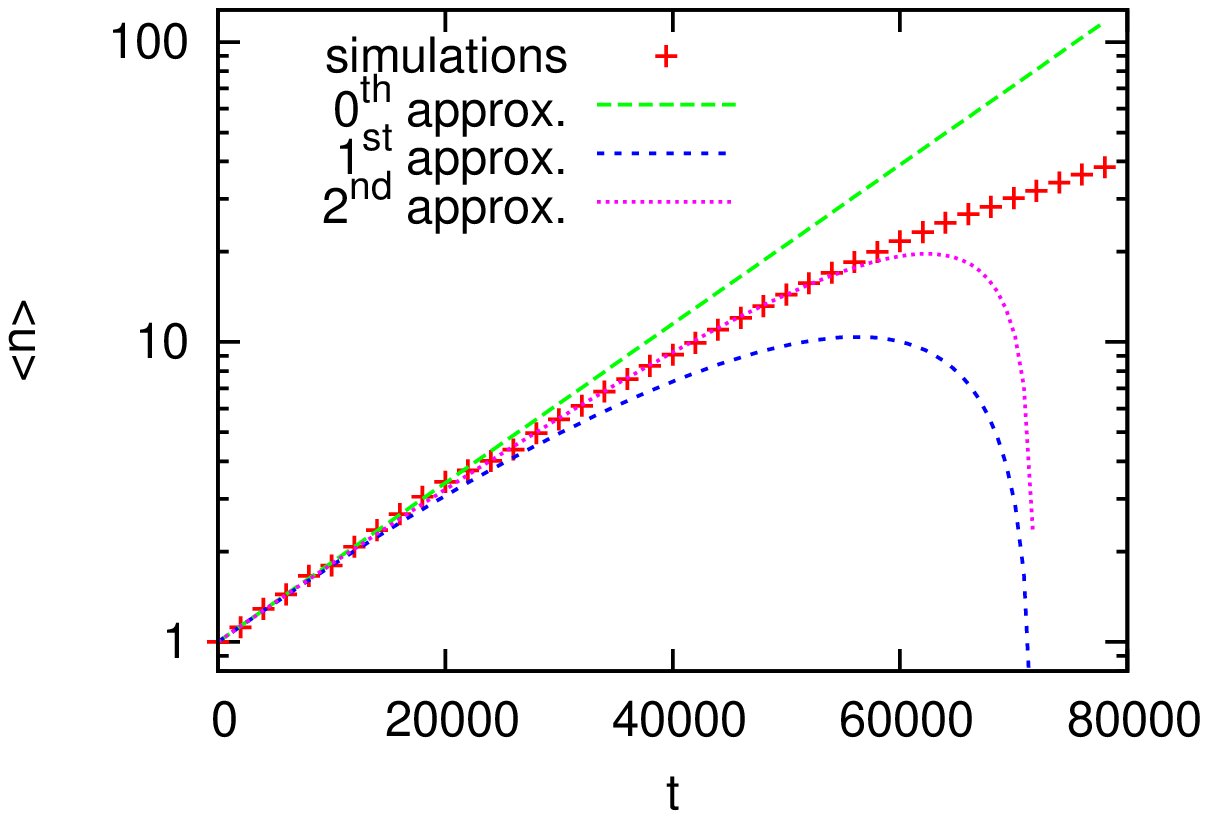} \\ 
(a) $\alpha=5 \cdot 10^{-4}$ & 
(b) $\alpha=7 \cdot 10^{-4}$ & 
(c) $\alpha= 10^{-3}$ \\ 
\end{tabular} 
\caption{(Color online) Chain topology, $N=128$, various levels of creativity $\alpha$. Comparison of approximations Eq. \eqref{eqn:n_avg_0}, Eq. \eqref{eqn:n_avg_1} and Eq. \eqref{eqn:n_avg_2}. The red crosses refer to the simulated data.}
\label{fig:comparison_of_approximations}
\end{figure*}

The comparison with the simulations (Fig. \ref{fig:comparison_of_approximations}) shows that the last approximation (Eq. \eqref{eqn:n_avg_2}) is better than the previous ones (Eq. \eqref{eqn:n_avg_0}, Eq. \eqref{eqn:n_avg_1}).

\section{CONCLUSIONS}

We have developed an analytical approach based on master equation that describes a model of paradigms evolution \cite{Bornholdt11} and confronted our results with the outcome of the work of Bornholdt et al. as well as with our numerical simulations. The outcome suggests, that the asymmetry between the paces of growth and decline of the dominant idea, observed in \cite{Bornholdt11}, is a generic property of the model and should be observed for any topology of interactions.

Our analytical methodology can be used to consider various topologies of interaction networks. The crucial parameter of the dynamics is the creativity of the agents, described by the $\alpha$ parameter. In the case when agents are almost non-innovative, the evolution consists of subsequent {\it periods of stagnation} (single paradigm present in the community) and {\it periods of expansion} (an innovative paradigm spreads across the community replacing the old one). The mean length of the {\it stagnation period} is equal to $\langle T_{stag} \rangle = 1/\alpha$, regardless of the interaction network topology. On the other hand, the mean length of the {\it expansion period} $\langle T \rangle$ strongly depends on the topology. If $\alpha \ll \langle T \rangle^{-1}$, the mean time between shifts of dominant paradigms can be approximated by $\langle T_{stag} + T \rangle \approx \langle T_{stag} \rangle = 1/\alpha$, which is the scaling observed in the simulated data by Bornholdt et al. \cite{Bornholdt11} (note that a different time scale was used in \cite{Bornholdt11}).

Our approach is mainly based on the approximation of two competing paradigms, which is justified if the level of creativity $\alpha$ is small enough, i.e. $\alpha < 1/\langle T \rangle$. For each type of interaction network topology this range has to be calculated separately. Five different topologies were taken into consideration: chain, complete graph, ER graphs, star and square lattice.

For the chain topology it was possible to find compact forms of the analytical solutions. The mean {\it expansion time} $\langle T\rangle$ scales with the system size $N$ like $N^2 \log N$ and during the {\it stage of expansion}, the mean size of the cluster of the new idea grows like a damped exponential function, see Eq. \eqref{eqn:chain_mean}. The analytical results are in agreement with the simulated data.

For the complete graph topology the proposed approach also results in a good agreement with the simulations, but compact forms of the functions describing the system evolution (Eq. \eqref{eqn:compl_P}, Eq. \eqref{eqn:compl_T_dist}) probably do not exist.

In the case of ER graphs topology, our approach underestimates the rate of the growth of the new idea cluster (Fig. \ref{fig:er_means}). The reason can be correlations between the node degree and the state of the agent located at that node, which were not taken into account.

Comparison of two variants of star topology brought interesting results. In the first variant, when we require that the innovation firstly appears in the central node, the rate of the expansion of the new idea is the fastest possible among all the topologies and the mean {\it expansion time} $\langle T \rangle$ (which is the shortest possible) scales like $N \log N$. However, if we remove this requirement and let the innovation appear in any node with equal probability, $\langle T \rangle$ grows to $N^3$ (higher than at a chain, where mean distance between nodes is much larger) and almost whole that time is taken by ``convincing'' the central agent to the new idea.

In the case of the square lattice, the method only qualitatively reproduces the results of the simulations (Fig. \ref{fig:sq_means}). The problem probably lies in the apparently too rough estimation of the shape of the cluster of the new idea as a circle.

For a higher level of creativity $\alpha$, when most of the time more than two ideas coexist, the dynamics of the system can be found starting from the results obtained for the case of lower levels of $\alpha$ and using the method similar to the perturbation method. This approach proved to be useful in the simplest case --- the chain topology. We were considering the function describing the mean number of agents following the {\it expanding} paradigm 1. It was found that the unperturbed function \eqref{eqn:n_avg_0}
\begin{equation}
\langle n(t) \rangle = e^{t/N^2}
\end{equation}
should be modified by two factors. The first one,
\begin{equation}
cos \left( \frac{\alpha}{N^3} t \right) < 1,
\end{equation}
describes ``attacking'' of the paradigm 1 by paradigms appearing after it. The second one,
\begin{equation}
1-\log\left|\cos\left(\frac{\alpha}{N^3} t\right)\right| - \frac{1}{4} \left(\frac{\alpha}{N^3}\right)^2 t^2 \approx 1 + \frac{1}{4} \left(\frac{\alpha}{N^3}\right)^2 t^2 > 1,
\end{equation}
describes ``attacking'' of the paradigms ``attacking'' paradigm 1. Both these terms, as expected, converge to 1 if the creativity of the agents $\alpha$ converges to 0.

Our analytical approach allows for a better understanding of the system dynamics described by the model \cite{Bornholdt11} and explains some of relationships previously observed in the simulated data. The proposed methodology can be used to analyze the dynamics of paradigms spreading in other networks \cite{Albert02}. It is especially interesting since the real networks of human contacts (including scientific collaboration networks) exhibit some nontrivial properties, such as scale-free behavior \cite{Newman01}. Investigations of such networks are planned in the future.


\begin{acknowledgments}
The authors acknowledge support from European COST Action MP0801 Physics of Competition and Conflicts, from Polish Ministry of Science Grant 578/N-COST/2009/0 and a special grant from Warsaw University of Technology.

G.S. is thankful for Tomasz Miller for fruitful discussions and remarks.
\end{acknowledgments}


\end{document}